\documentclass[12pt,english,5p,times,sort&compress,pdftex]{elsarticle}
\usepackage[T1]{fontenc}
\usepackage[latin9]{inputenc}
\usepackage{color}
\usepackage{array}
\usepackage{booktabs}
\usepackage{textcomp}
\usepackage{amsmath}
\usepackage{graphicx}
\usepackage{amssymb}

\makeatletter

\newcommand{\noun}[1]{\textsc{#1}}
\providecommand{\tabularnewline}{\\}

\journal{Cryogenics}
\renewcommand{\@makecaption}[2]{{\centering   \vskip\abovecaptionskip \bfseries #1:} #2}
 
\renewcommand*{\l@figure}[2]{ %
\setlength\@tempdima{2.3em} %
\noindent\hspace*{1.5em}Figure #1\hfil\newline\newline }

\renewcommand*{\l@table}[2]{ %
\setlength\@tempdima{2.3em} %
\noindent\hspace*{1.5em}Table #1\hfil\newline\newline }

\makeatother

\usepackage{babel}

\begin{document}

\title{Two-stage high frequency pulse tube cooler\\
 for refrigeration at 25 K }

\author{M. Dietrich\corref{md}}

\ead{marc.dietrich@ap.physik.uni-giessen.de}

\author{G. Thummes\corref{gt}}

\address{Institute of Applied Physics, University of Giessen, D-35392 Giessen,
Germany\\
and\\
TransMIT-Centre for Adaptive Cryotechnology and Sensors, D-35392
Giessen, Germany}
\begin{abstract}
A two-stage Stirling-type U-shape pulse tube cryocooler driven by
a 10 kW-class linear compressor was designed, built and tested. A
special feature of the cold head is the absence of a heat exchanger
at the cold end of the first stage, since the intended application
requires no cooling power at this intermediate temperature. Simulations
where done using \noun{Sage}-software to find optimum operating conditions
and cold head geometry. Flow-impedance matching was required to connect
the compressor designed for 60 Hz operation to the 40 Hz cold head.
A cooling power of 12.9 W at 25 K with an electrical input power of
4.6 kW has been achieved up to now. The lowest temperature reached
is 13.7 K.\end{abstract}

\cortext[md]{Corresponding author: Tel.: +49 641 9933462.}

\maketitle

\section{Introduction }

The recent development efforts for High-Temperature-Superconducting
(HTS) applications like transformers, generators or motors create
a demand on large scale cryocoolers\citet{Gromol2004}. While Gifford-McMahon
(GM)- and oil-lubricated Stirling-coolers are presently the prevalent
choice in HTS power applications\citet{Gromol2004,Hasselt2004,Hezelton2004},
pulse tube coolers (PTC) promise a higher reliability and longer maintenance
intervals. Large scale PTCs of the GM-type have been used for low
temperature cryocooling since several years now, mostly in the 4 K
region. While it is possible with this type of cooler to achieve large
cooling powers in the 30 K region, they also suffer from relative
short maintenance intervals of the compressor. Also the efficiency
of the compressor/rotary-valve combination is quite low. Therefore,
high frequency Stirling-type PTCs with oil-free linear compressors
became more attractive since they promise better cooling performance
than GM-type coolers as well as longer maintenance intervals. During
the last years, several large scale single-stage Stirling type PTCs
for operating temperatures between 60 and 80 K have been successfully
developed\citet{Zia2007,Potratz2008,Ercolani2008}. 

The work presented in the present paper is part of our ongoing efforts
to develop a Stirling type pulse tube cryocooler for neon recondensation
near 25 K\citet{Gromoll2006,Dietrich2007,Sun2009}. Among others,
this temperature is required to operate a BSCCO-based 400 kW HTS motor
developed at Siemens for ship propulsion\citet{Hasselt2004}. The
low temperature allows for higher current densities inside the rotor
coils and thus for a more compact design. During our high-power PTC
development some limitations arose from a new kind of regenerator
streaming that results from a poor thermal communication of the regenerator
gas in transverse direction, as is also described in\citet{So2006}.
This streaming starts when the cold end temperature falls below a
critical value and/or when a critical acoustic power (or mass flow)
is exceeded. Because of an excess heat transfer to the cold end, the
streaming significantly degrates the cooling performance. Up to now,
the critical operating conditions for the onset of such a streaming
cannot be quantitatively predicted\citet{So2006}. Possible solutions
like enhancing the radial thermal conductivity of the regenerator
matrix imply other losses caused by heat transfer inside the regenerator\citet{Dietrich2007}.
So far the only solution to avoid decreasing the regenerator performance
is to lower the amount of acoustic power.

A two-stage, gas-coupled cryocooler provides a way to solve these
problems. First, a part of the acoustic power is directed into the
first stage, thus reducing the acoustic power in the second stage.
Second, the small acoustic power in the second stage allows a reduced
regenerator diameter, which enhances the radial thermal communication
of the working gas. Both measures reduce the risk of that kind of
regenerator streaming. We therefore decided to develop a new, two-stage
gas-coupled cold head, to minimize the possibility of regenerator
streaming.

\section{Modelling of the cold head}

\subsection{Design considerations\label{sub:Design-considerations}}

The cold head was modelled by maintaining two constraints: Firstly,
to reduce the possibility of regenerator streaming, the acoustic power
inside the regenerator should not exceed a critical value. Because
theoretical predictions of the onset of such streaming are still under
development\citet{So2006}, we relied on our experiences from former
designs\citet{Dietrich2007} and limited the acoustic power $\dot{W}_{pV}$
to 2 kW. It was clear that the target cooling power $\dot{Q}_{total}$
of 60 W @ 25 K could not be obtained with such a low power, so the
only solution was to build several identical cold heads for parallel
operation on one compressor. The necessary number \textit{n} of parallel
cold heads was estimated from\begin{equation}
\dot{Q}_{total}=n\,\dot{Q}_{c}=n\,\eta\,\frac{T_{c}}{T_{h}-T_{c}}\,\dot{W}_{pV},\label{eq:1}\end{equation}
where $\dot{Q}_{c}$ is the cooling power of one cold head at temperature
$T_{c}$, $T_{h}$ the ambient temperature, and $\eta$ the efficiency
of the cooler. From the pulse tube design of Nguyen et al.\citet{Nguyen2004},
we estimated the target efficiency at $T_{c}=$ 25 K and $T_{h}=$
300 K to $\eta=$ 10\% of Carnot relative to pV-power. Then Eq.~\ref{eq:1}
gives a rounded-up number of \textit{n} = 4 cold heads.

Secondly, an important factor is that the required pV-power of 8 kW
should be delivered by the available 10 kW compressor (Qdrive model
2S297W) which was designed for an optimum working frequency of 60
Hz. However, consideration of the regenerator losses at low temperatures
and high frequency revealed that the cold heads must be driven at
a lower frequency in order to reach high efficiency (see next section).
Another problem of using high frequencies is the proper flow straightening
at high gas velocities at low gas displacement amplitudes. Since the
pV-power scales roughly linearly with the working frequency, the complete
cooler could not be tested with the present compressor. We therefore
decided to test only one of the four parallel cold heads and to connect
a more appropriate compressor at a later date.

\subsection{Cooler configuration}

While two-stage PTCs with thermally coupled stages are able to reach
low temperatures\citet{Jiang2004,Tang2005}, there are difficulties
to transfer heat over large cross-sections, which are required in
high-frequency, high power PTCs\citet{Sun2009}. Another problem is
the large dead volume at the cold end of a regenerator with uniform
cross-section that is caused by flow straighteners and the \textcolor{black}{connection}
to the pulse tube, which has to be charged by the mass flow from the
warm end. Compared to a gas-coupled cooler, where the cross-section
is staged, this causes an increase in regenerator losses. While this
may be solved by use of a conical regenerator, a gas-coupled two-stage
cold head was favoured for practical reasons.

Another advantage of a gas-coupled systems arises, when no cooling
power is required at the first stage, as is the case with the intended
HTS-motor cooling. In usual two-stage pulse tube systems, a solid
flow channel connects the first-stage pulse tube to the joint between
the two regenerators, where some kind of flow separator divides the
gas flow from the first-stage regenerator. Normally, the connecting
channel is small in order to reduce void volume and to allow the cold
gas from the first stage to exchange heat with the solid. From a thermodynamic
point of view, each heat flow $\dot{Q}$ across a temperature difference
$\Delta T$ at a mean temperature $T$ produces irreversible entropy:\begin{equation}
S_{irr}\approx\frac{Q<\Delta T>}{<T^{2}>},\label{eq:2}\end{equation}
which leads to a decrease in exergy $\Xi=H-T_{h}S$\citet{Kittel2005}.
The loss from eq.~(2) can be avoided by eliminating such heat transfer
areas and by thermally decoupling the first stage pulse and regenerator,
e.g. by the use of a connecting line and flow straighteners with low
thermal conductance. In such a system, the first stage will be purely
{}``enthalpy coupled'' avoiding irreversible entropy production
by heat transfer. 

At the beginning of the design, a basic \noun{Sage\citet{Gedeon1994}}
model without compressor was set up to find the optimum operating
conditions. Using \noun{Sage,} the regenerator mesh of the first stage
was fixed to stainless steel with mesh \#400 (25 \textmu{}m wire diameter)
and the second stage was filled with lead spheres. The phase shifters
consist of single inertance lines and a second-inlet at the second
stage. The filling pressure was set to 2.5 MPa and the maximum pressure
amplitude to 0.4 MPa as this is the limit of the 2S297W compressor.
Simulations were performed to analyze the effect of operating frequency
and input power. Fig.%
\begin{figure}
\begin{centering}
\includegraphics[width=1\columnwidth]{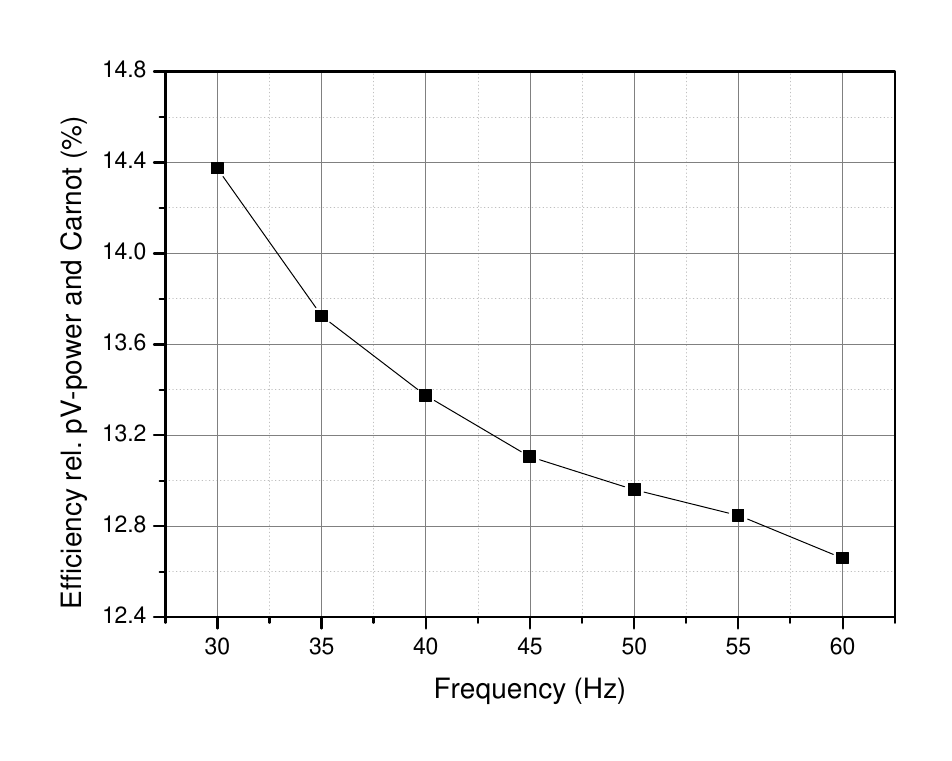}
\par\end{centering}

\caption{Calculated efficiency relative to pV-power and Carnot in a frequency
range from 30 to 60 Hz with a pV-power of 2 kW.\label{fig:eta_vs_freq_sim}}

\end{figure}
\ref{fig:eta_vs_freq_sim} shows the cold head efficiency in a frequency
range between 30 and 60 Hz with a pV-power of 2 kW. In each simulation,
the lengths and diameters of the regenerators, pulse tubes and phase
shifters where optimized to give maximum cooling power at 25 K. Also
the lead sphere diameter was optimized, while the geometry of the
heat exchangers was not varied in order to speed up the optimization
process. It can be seen from Fig.\ref{fig:eta_vs_freq_sim} that the
efficiency increases when going to lower frequencies. Most of the
losses arise from regenerator inefficiency due to friction loss and
small thermal penetration depth. While operation at a lower frequency
seems to be better, one must also take into account that the compressor
must be able to deliver the required acoustic power at low frequency.
Therefore, as a compromise we have chosen a working frequency of 40~Hz.

\section{System setup}

Fig.%
\begin{figure*}
\centering{}\includegraphics[width=0.8\textwidth]{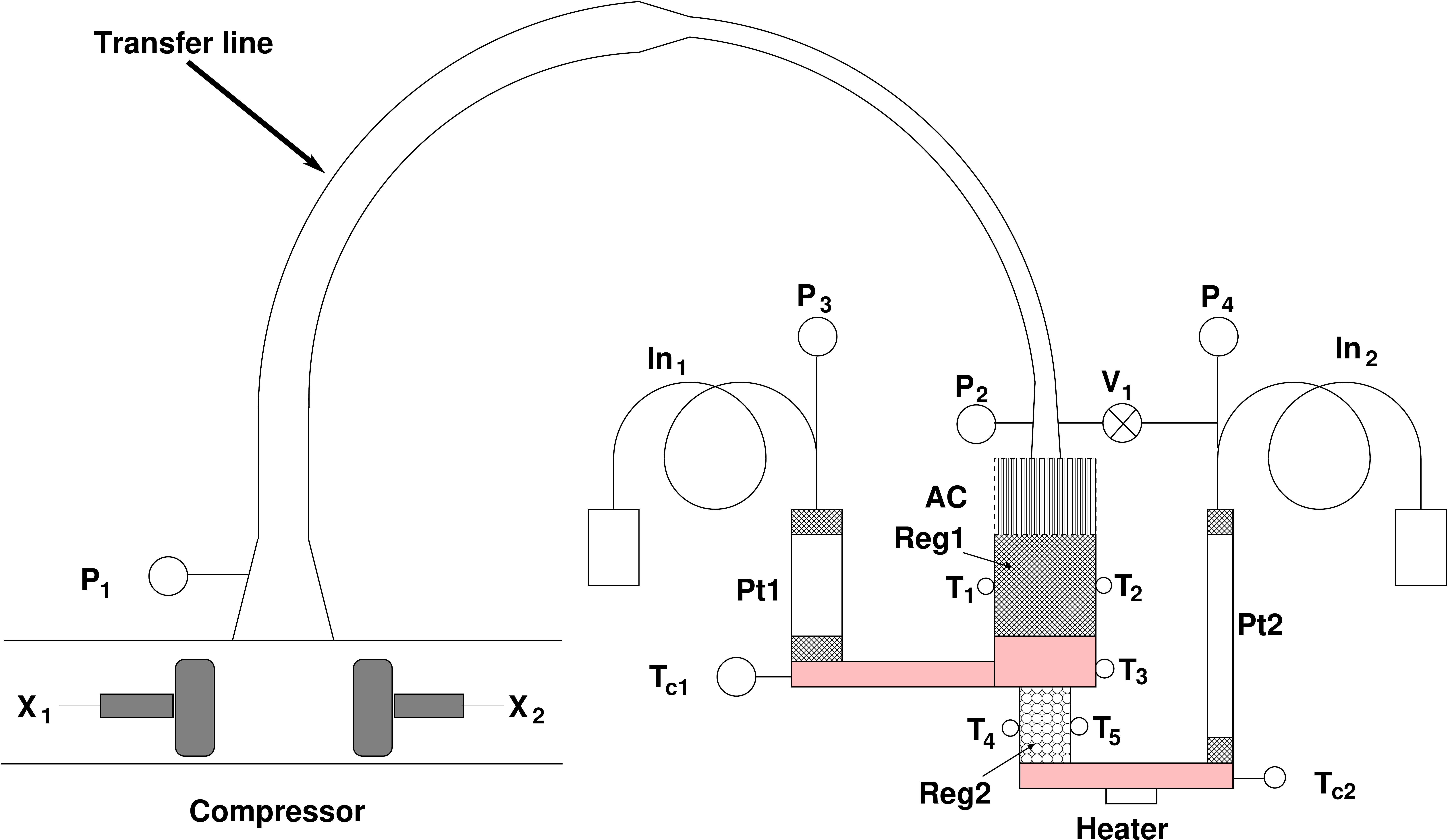}\caption{Schematic view of the system setup. $X_{1}$; $X_{2}$: position sensors,
$P_{1-4}$: pressure sensors, $T_{1-5}$; $T_{c1}$; $T_{c2}$: temperature
sensors, $AC$: aftercooler, $Reg1$; $Reg2$: regenerators, $Pt1$;
$Pt2$: pulse tubes, $In_{1}$; $In_{2}$: inertance and buffer, $V_{1}$:
second-inlet\label{fig:Schematic}.}

\end{figure*}
\ref{fig:Schematic} shows the system setup. The data acquisition
setup is similar to that described in\citet{Dietrich2007}. The cold
head is connected to the compressor via a transfer line, which consists
of two tubes with different diameters connected in series. As described
in section 4, the transfer line also acts as an impedance matching
device. The aftercooler is of shell and tube type. The first stage
regenerator is filled with \#325/28 \textmu{}m and \#400/25 \textmu{}m
stainless steel screens. The flow distributor between the two regenerators
is filled with an alternating stack of \#80 and \#400 screens. The
second stage regenerator is filled with lead spheres of about 100
\textmu{}m diameter.\textcolor{red}{{} }The lengths (diameters) of the
first and second stage pulse tubes are 70 mm (26 mm), and 107 mm (16
mm), respectively. Like the transfer line, the inertance tubes also
consist of a series of two tubes with different diameters. 

The linear compressor is equipped with position sensors $X_{1}$ and
$X_{2}$. Together with pressure sensor $P_{1}$ located in the compression
space, it is possible to measure the acoustic power delivered by the
compressor in real-time. Pressure sensors are also located at the
entrance to the aftercooler ($P_{2}$) and at the hot ends of the
pulse tubes ($P_{3}$ and $P_{4}$). Temperature sensors are located
at the cold ends of the pulse tubes ( $T_{c1}$ and $T_{c2}$) and
pairwise with a 180\textdegree{} azimuthal shift at the mid positions
of the regenerators ($T_{1}$, $T_{2}$ and $T_{4}$, $T_{5}$) to
detect a possible flow inhomogeneity. There is also an additional
sensor installed at the flow distributor ($T_{3}$). All temperature
sensors except $T_{c2}$ are of PT-100 type. $T_{c2}$ is a cernox
sensor. An electrical heater is installed at the bottom of the second
stage heat exchanger to enable the measurement of cooling power. The
second stage is equipped with an adjustable second-inlet ($V_{1}$).

\section{Properties of the impedance matching line}

After finishing the basic design of the cold head, the compressor
model was added to the \noun{Sage} model. The standard way to adapt
the {}``60 Hz'' compressor to the cold head and operate them at
40 Hz would have been to add a rather large void volume in between
them\citet{Martin2002}. However, it was found from our simulations
that the required amplitude of the pressure wave cannot be achieved
by the use of a void volume. Instead, we have chosen a transfer line,
which acts simultaneously as an impedance matching device and as pressure
wave amplifier. The length of the line is 6 m, which roughly corresponds
to one fourth of the acoustic wave length of helium at 40 Hz. Fig.%
\begin{figure}
\begin{centering}
\includegraphics[width=1\columnwidth]{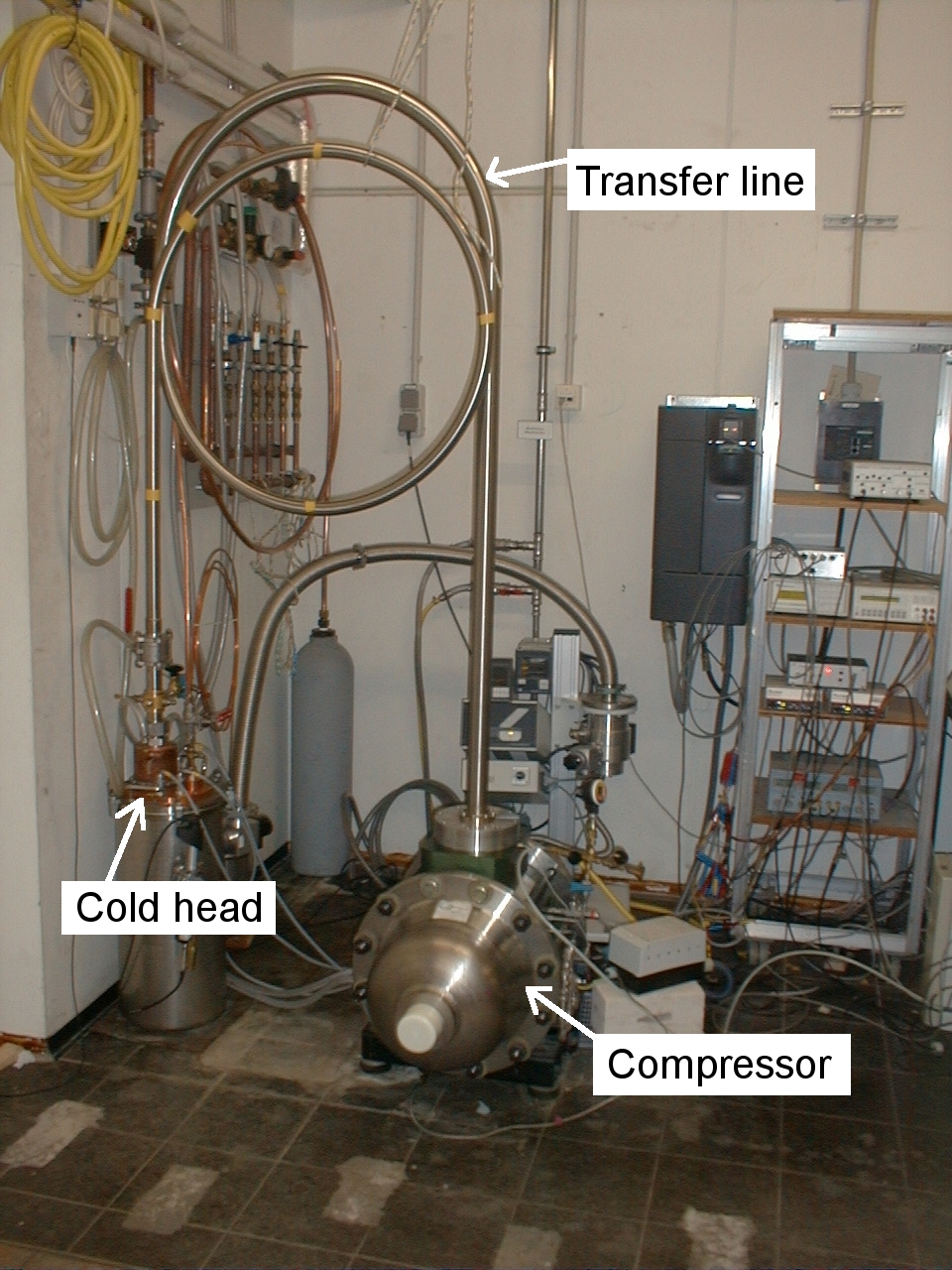}
\par\end{centering}

\caption{Photograph of the pulse tube cooler; the cold head is connected to
the linear compressor via the 6 m transfer line.\label{fig:pic_of_coldhead}}

\end{figure}
\ref{fig:pic_of_coldhead} shows a photograph of the setup. In order
to reduce losses, the transfer line consists of two parts with different
diameters of 43 mm and 30 mm. Fig.%
\begin{figure}
\begin{centering}
\includegraphics[width=1\columnwidth]{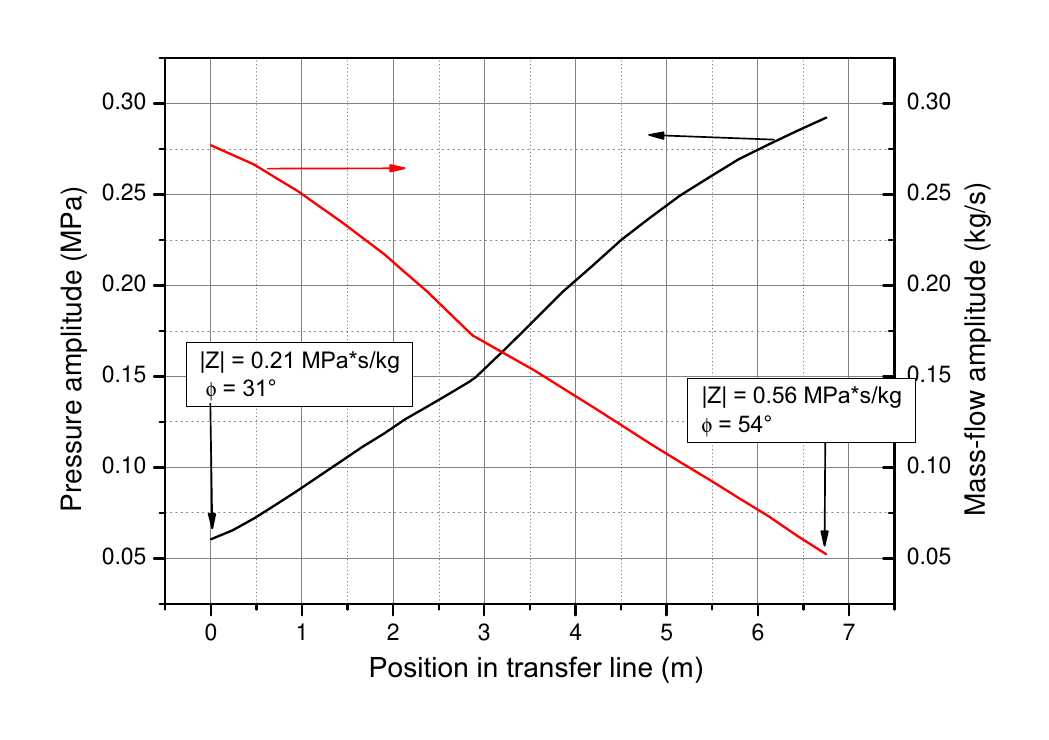}
\par\end{centering}

\caption{Calculated pressure and mass flow amplitude at different transfer
line positions, acoustic power at the compressor side of the transfer
line: $\dot{W}_{pV,in}=$1.9 kW.\label{fig:p_and_mdot_vs_tl_pos}}

\end{figure}
\ref{fig:p_and_mdot_vs_tl_pos} shows the calculated pressure and
mass flow amplitudes in the transfer line. The input and output impedance
are also given in the graph. It is seen that the transfer line amplifies
the pressure amplitude of 60 kPa at the inlet to 290 kPa at the outlet
(cold head side). The amplification of the pressure amplitude is accompanied
by a 38\% loss of acoustic power between compressor and cold head,
as seen from Fig.%
\begin{figure}
\begin{centering}
\includegraphics[width=1\columnwidth]{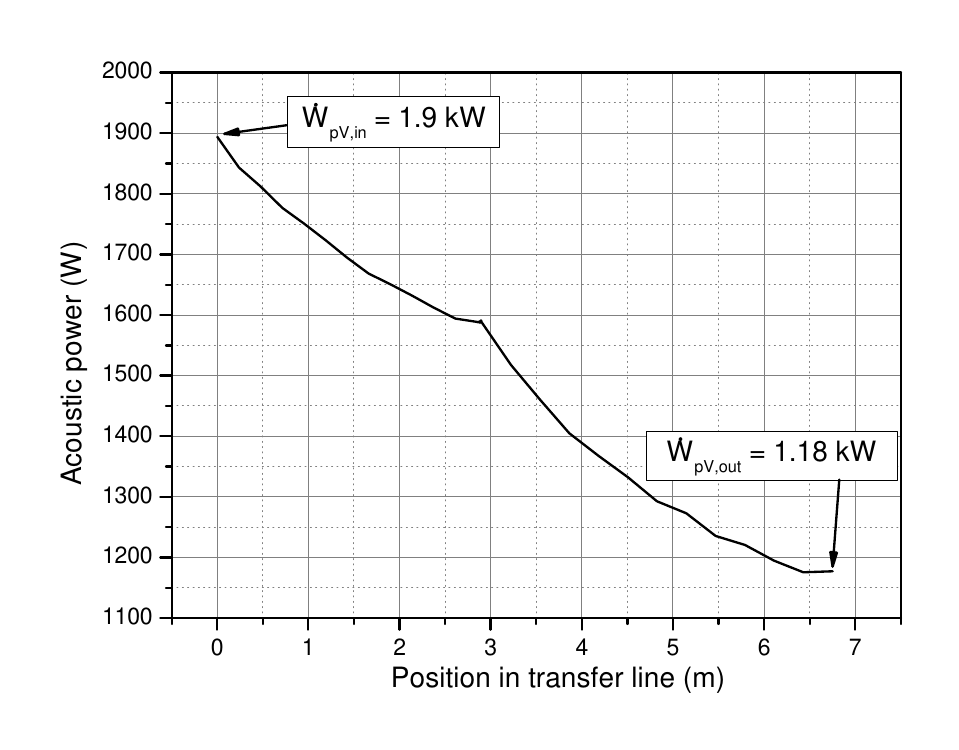}
\par\end{centering}

\caption{Decrease of the acoustic power $\dot{W}_{pV}$ in the transfer line.\label{fig:pv-power_in_tl}}

\end{figure}
\ref{fig:pv-power_in_tl}. During the validation of the computer model,
significant differences were found between the measured and the calculated
pressure phase shifts at both ends of the transfer line, which also
affected the pV-power. We attribute this to the bended structure of
the transfer line, which cannot be considered in the model. Upon lengthening
the transfer line in the model by 10\% a better matching with the
experimental results has been accomplished. The difference in pressure
amplitudes between the simulation and the measurement went down from
35 kPa to 10 kPa (see also Fig. \ref{fig:p_vs_pV}), while the pressure
phase angle error decreased from over 20\textdegree{} to less than
5\textdegree{} inside the compressor and from 5\textdegree{} to less
than 2\textdegree{} at the end of the transfer line.

\section{Experimental results and discussion}

During the optimization process, several modifications to the cold
head were made. In the first runs, a flow maldistribution in the first
stage pulse tube was found and fixed by adding more flow straightener
mesh at the cold end. Later optimizations concentrated on the flow
distributor between the two regenerators. Simultaneously with the
measurements, the simulation model was constantly improved to reflect
the exact construction of the cold head. All measurement were performed
at 2.5 MPa filling pressure and 40 Hz operating frequency.

Fig.%
\begin{figure}
\begin{centering}
\includegraphics[width=1\columnwidth]{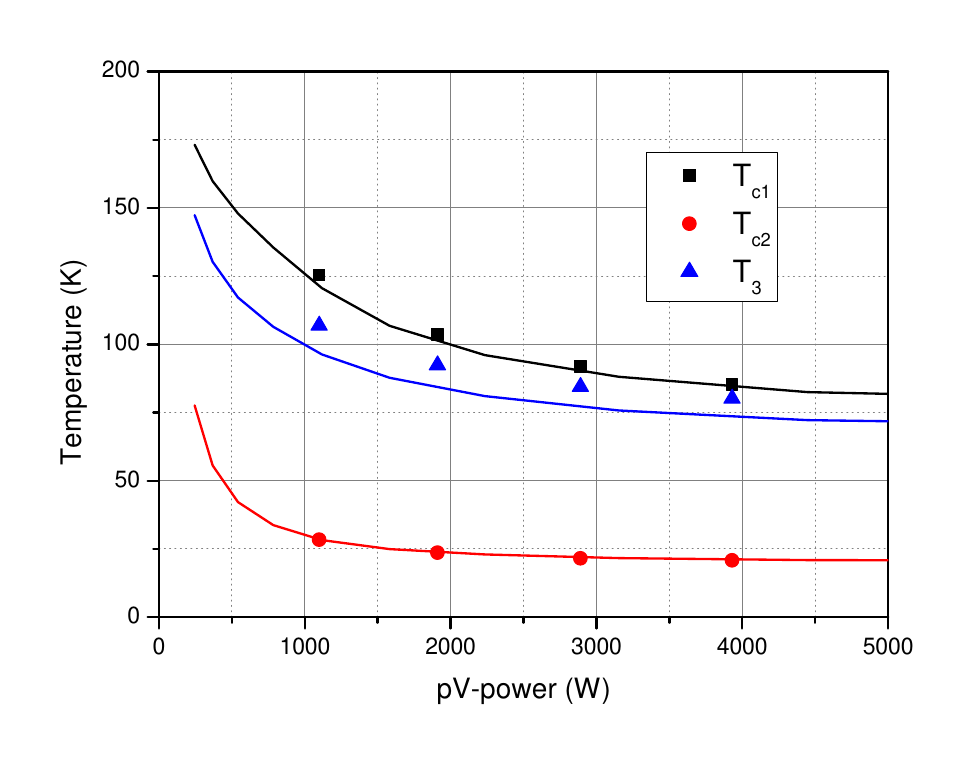}
\par\end{centering}

\caption{Measured temperatures $T_{c1}$, $T_{c2}$, and $T_{3}$ in inertance
mode (symbols) versus pV-power and comparison to simulation results
(solid lines) .\label{fig:T_sim_vs_exp}}

\end{figure}
\ref{fig:T_sim_vs_exp} shows the measured temperatures at the cold
ends of the first and second stage pulse tubes and at the flow distributor
between the regenerators as function of pV-power in inertance mode
(second-inlet closed) compared to the corresponding simulation results.
There is a good agreement for the cold end temperatures $T_{c1}$
and $T_{c2}$, but a slight mismatch for the temperature $T_{3}$,
which we attribute to some flow inhomogeneities inside the flow distributor. 

Fig.%
\begin{figure}
\begin{centering}
\includegraphics[width=1\columnwidth]{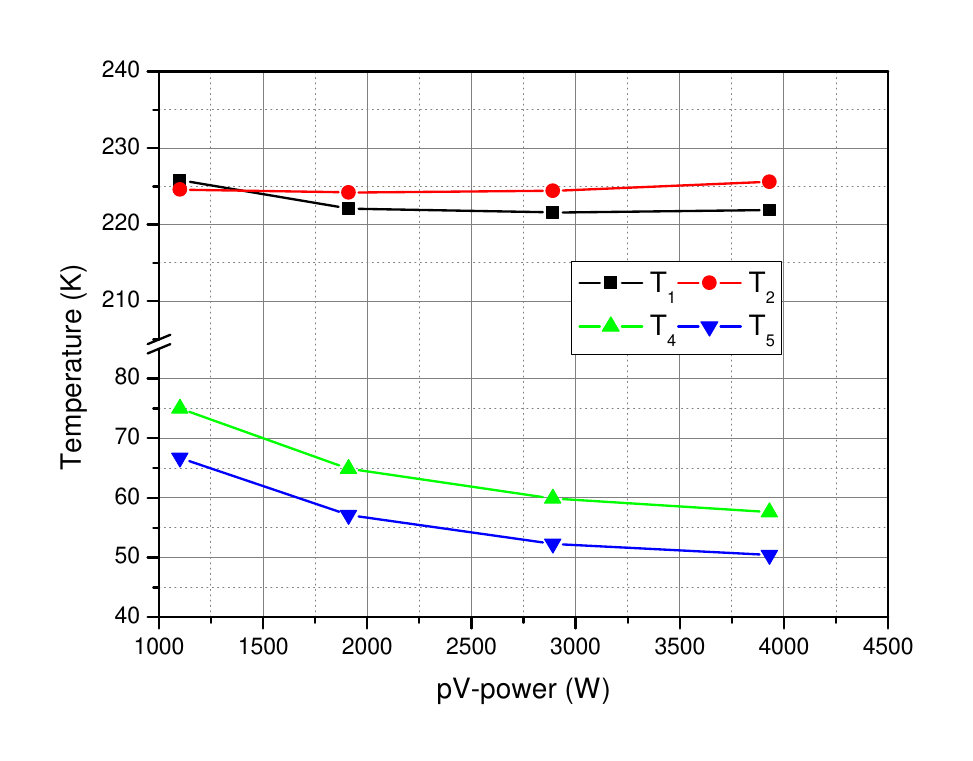}
\par\end{centering}

\caption{Measured temperatures in the middle of the regenerators versus pV-power.\label{fig:Treg_vs_pV}}

\end{figure}
\ref{fig:Treg_vs_pV} shows the temperature sensor readings $T_{1}$,
$T_{2}$ and $T_{4}$, $T_{5}$ (see Fig. \ref{fig:Schematic}) at
the middle positions of the two regenerators as function of the pV-power.
There is a slight difference between $T_{1}$ and $T_{2}$ at higher
pV-powers in the first regenerator, while in the smaller second regenerator
a constant offset exists, which is attributed to an inexact positioning
of the sensors. 

Fig.%
\begin{figure}
\begin{centering}
\includegraphics[width=1\columnwidth]{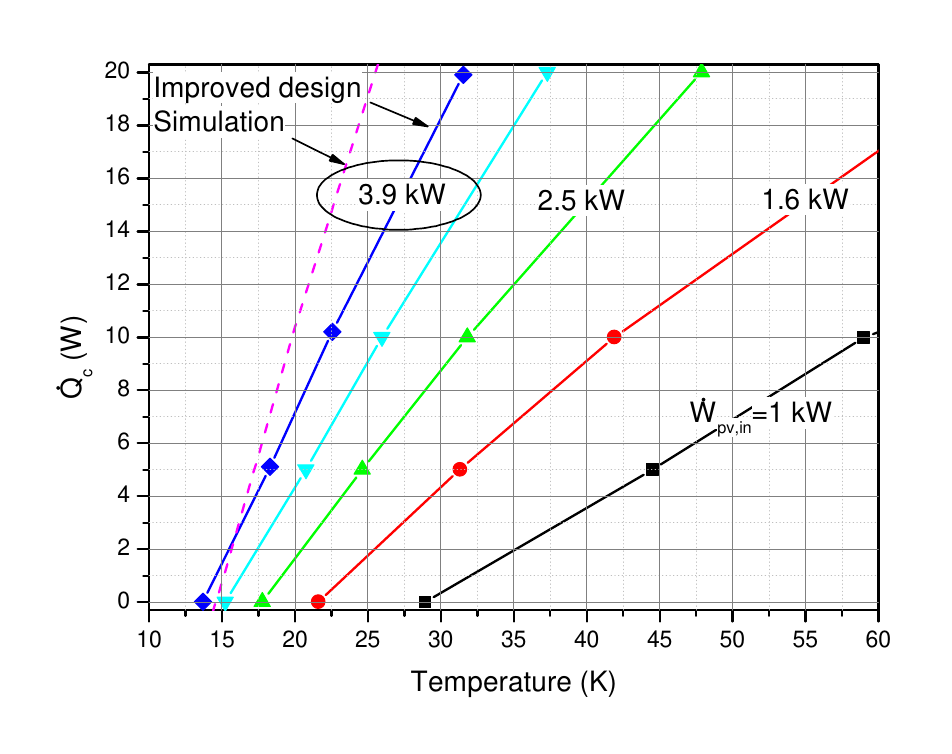}
\par\end{centering}

\caption{The dashed line shows a result from simulation with $\dot{W}_{pV,in}=$~3.9~kW.
The solid lines to the experimental data are guide for the eyes.\label{fig:cp_vs_pV}}

\end{figure}
\ref{fig:cp_vs_pV} shows the cooling powers for various pV-powers
in second-inlet mode. Before these measurements the second-inlet was
adjusted for minimum no-load temperature. After the initial measurements
took place, a modified cold head was manufactured, which had less
dead volumes than the original one. As seen from Fig. \ref{fig:cp_vs_pV},
this improved version is able to produce cooling power of 12.9~W
at 25~K with 3.9~kW of pV-power, corresponding to 2.4~kW acoustic
power at the inlet to the cold head, because of the loss in the transfer
line (see Fig. \ref{fig:pv-power_in_tl}). The electric input power
to the compressor is then 4.7~kW at 90\% of the full stroke. The
lowest temperature of 13.7~K is reached in the same measurement with
a slope of the load line of 1.15 W/K. Given the loss in the transfer
line, this corresponds to an efficiency of 5.6\% relative to the acoustic
power at the cold head and Carnot, which is only half of the target
efficiency estimated in section \ref{sub:Design-considerations}.
The dashed line shows the simulated result with 3.9 kW of pV-power,
which gives a cooling power of 19 W at 25 K at a much larger slope
of 1.75 W/K and a slightly higher no-load temperature. While the slope
can be attributed to the smaller pressure drop in the model compared
to the experiment, a higher no load temperature points to some inaccuracies
of the model parameters which are not yet identified.

Fig.%
\begin{figure}
\begin{centering}
\includegraphics[width=1\columnwidth]{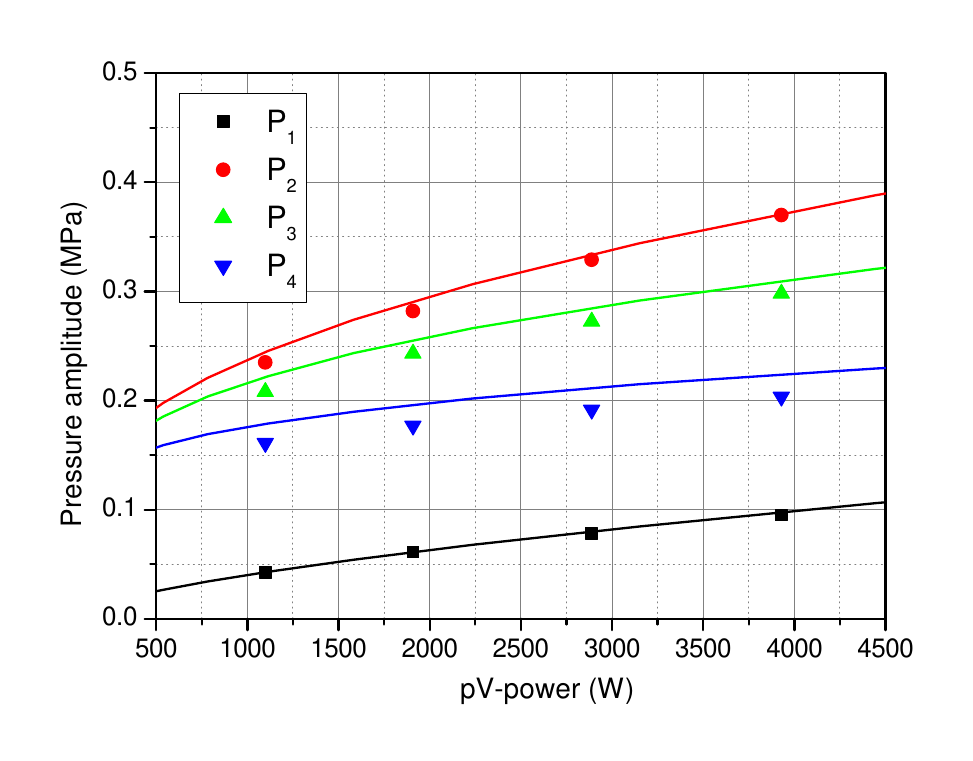}
\par\end{centering}

\caption{Pressure amplitudes in inertance mode at different pV-powers from
measurement (symbols) and from simulation (solid lines).\label{fig:p_vs_pV}}

\end{figure}
\ref{fig:p_vs_pV} compares the measured and simulated pressure amplitudes
at different pV-powers in inertance mode (second-inlet closed). While
for the pressure amplitudes in the compressor and at the hot end of
the regenerator the agreement between measurement and simulation is
good, the pressure amplitudes in the pulse tubes are lower in the
experiment than in the simulation. In case of the second pulse tube
the difference amounts for 10\%. This is significant as the enthalpy
flow in the pulse tube roughly scales with the square of the pressure
amplitude. Therefore, compared to the experiment the calculated enthalpy
flow in the second pulse tube should be larger by 20\%. Table%
\begin{table}
\caption{Calculated enthalpy flows in the second stage regenerator and pulse
tube and net cooling power with second-inlet closed at a pV-power
$\dot{W}_{pV,in}=$ 3.9 kW and $T_{c2}=$ 25 K. The values in parenthesis
show the simulation results with additional pressure drop.\label{tab:enhalpy-flows}}

\centering{}{\small }\begin{tabular}{>{\centering}m{0.3\columnwidth}>{\centering}m{0.3\columnwidth}>{\centering}m{0.2\columnwidth}}
\toprule 
\addlinespace
 & {\small Enthalpy flow}{\small \par}

{\small (W)} & {\small Net cooling power}{\small \par}

{\small (W)}\tabularnewline\addlinespace
\midrule
\addlinespace
{\small Regenerator 2} & {\small 25.2 (20)} & \tabularnewline\addlinespace
\midrule 
\addlinespace
{\small Pulse tube 2} & {\small 31.6 (20.8)} & {\small 6.4 (0.8)}\tabularnewline\addlinespace
\bottomrule
\end{tabular}
\end{table}
\ref{tab:enhalpy-flows} shows the calculated enthalpy flows including
conduction losses in the second stage regenerator and pulse tube with
second-inlet closed. A modified model which has the additional pressure
drop simulated by adding some local loss in the second stage cold
heat exchanger was made. These results are shown in parentheses in
table \ref{tab:enhalpy-flows}. Due to the additional pressure drop
the enthalpy flow in the regenerator decreases by about 5 W (21\%)
and the enthalpy flow in the pulse tube decreases by about 11 W (33\%).
The difference in enthalpy flows between the pulse tube and the second
regenerator gives the net cooling power $\dot{Q}_{c}$, which is about
6 W larger without the additional pressure drop. The reason for the
larger measured pressure drop is not yet identified, but is hoped
to be fixed in the next cold head version.

\section{Conclusions}

We have developed a Stirling-type two-stage pulse tube cryocooler
system for neon recondensation near 25 K. In order to avoid problems
with streaming instabilities in regenerators with large cross-section
area, the system design is based on four smaller PTC cold heads to
be operated in parallel on a single linear compressor. From experiments
on one of the PTC cold heads in combination with a 10 kW-class compressor
it is shown that the two-stage design successfully suppresses the
onset of internal regenerator streaming. It is also confirmed by the
measurements that a solely \textquotedblleft{}enthalpy-coupled\textquotedblright{}
configuration of the stages without cold heat exchanger at the first
stage is feasible, thereby reducing the losses in the first stage.

For matching the compressor to the cold head at 40 Hz operation, a
6 m long transfer line has been used that amplifies the pressure amplitude
but also reduces the available pV-power at the cold head to about
62\% of that at the compressor. A no-load temperature of 13.7 K and
a cooling power of 12.9 W at 25 K have been achieved at an electrical
input power of 4.7 kW, which corresponds to 3.9 kW of acoustic power
at the compressor and to 2.4 kW of acoustic power at the cold head.
The corresponding efficiency at 25~K is 5.6\% relative to Carnot
and pV-power at the cold head. 

The calculated cooling power from simulation amounts to 19~W at 25~K
with $W_{pV,in}=$ 3.9~kW as compared to the measured 12.9~W. The
lower experimental cooling power can be attributed to a rather large
pressure drop in the second stage, as found from a comparison between
the measured and calculated pressure amplitudes. It is hoped that
in a revised version of the cold head the pressure drop can be reduced
and thus the cooling power increased to near 20~W at 25~K.

Then, by using a larger linear compressor, designed for 40 Hz operation
and capable of delivering about 10~kW of pV-power, four parallel
cold heads should be able to deliver 80~W of cooling power at 25
K for neon recondensation.

\section*{Acknowledgement}

This work is supported by the German Ministry of Economics and Technology
(BMWi) under grant no. 03SX221A. We thank the Siemens AG (Erlangen,
Germany) for providing the linear compressor.

\newpage{}\bibliographystyle{cryogenics}
\addcontentsline{toc}{section}{\refname}\nocite{*}
\bibliography{cryo5}

\newpage{}
\end{document}